\documentclass[aps,prl,twocolumn,showpacs,superscriptaddress,groupedaddress]{revtex4}  
\usepackage{graphicx}  
\usepackage{dcolumn}   
\usepackage{bm}        
\usepackage{amssymb}   

\begin{document}

\widetext
\leftline{Version 1.00 as of \today}

\title{Turbulent momentum pinch of diamagnetic flows in a tokamak}
\author{Jungpyo Lee} \affiliation{Plasma Science and Fusion Center, MIT, Cambridge, USA}
\author{Felix I. Parra} \affiliation{Plasma Science and Fusion Center, MIT, Cambridge, USA}
\author{Michael Barnes} \affiliation{Plasma Science and Fusion Center, MIT, Cambridge, USA}
\date{\today}

\begin{abstract}
The ion toroidal rotation in a tokamak consists of an $E\times B$ flow due to the radial electric field and a diamagnetic flow due to the radial pressure gradient. The turbulent pinch of toroidal angular momentum due to the Coriolis force studied in previous work is only applicable to the $E\times B$ flow. In this Letter, the momentum pinch for the rotation generated by the radial pressure gradient is calculated and is compared with the Coriolis pinch. This distinction is important for subsonic flows or the flow in the pedestal where the two types of flows are similar in size and opposite in direction. In the edge, the different pinches due to the opposite rotations can result in intrinsic momentum transport that gives significant rotation peaking. 
\end{abstract}

\maketitle


\emph{Introduction.}  Ion toroidal rotation can stabilize MHD resistive wall modes \cite{Bondeson:PRL1994, Strait:PRL1995} and increases the confinement time in tokamak plasmas \cite{Barnes:PRL2011shear,Highcock:PRL2010,Parra:PRL2011}. Consequently, external momentum sources are frequently used to drive rotation in current tokamaks. However, for larger volume, high density plasmas such as those expected in ITER, the available external sources are insufficient to produce significant toroidal rotation. Thus, we can only rely on rotation that is ``intrinsically" generated by the radial transport of the toroidal angular momentum due to microturbulence and by coupling to the rotation at the plasma-wall boundary. Radial turbulent transport redistributes the toroidal momentum within a tokamak.

The intrinsic ion toroidal rotation has been measured in many tokamaks \cite{Rice:NF2007}, and the size of the flow is much below the ion thermal velocity (Mach number $\sim 0.1- 0.2$). The subsonic level of the flow is in accordance with theoretical estimations \cite{Sugama:PPCF2011,Parra:PRL2012}. Due to the symmetry properties of the turbulence in an up-down symmetric tokamak, the rotation intrinsically generated by momentum redistribution is subsonic when the rotation at the edge is subsonic, as it usually is. 
The reason is that for sonic flow a symmetry of the turbulence that gives vanishing toroidal angular momentum transport, and hence momentum redistribution, can only be broken by pre-existing nonzero toroidal flow or radial flow shear of the flow, giving turbulent momentum pinch and diffusion, respectively \cite{Peeters:NF2011}. 

The momentum pinch is inward momentum flux that depends on the flow size \cite{Peeters:PRL2007,Hahm:POP2007}. For small flows, this momentum flux is simply proportional to the flow. It is not just convective transport due to the radial particle pinch carrying momentum, but also momentum flux caused by the presence of non-zero rotation even in the absence of the particle pinch.


In previous work about the momentum pinch \cite{Hahm:POP2007,Peeters:PRL2007, Peeters:POP2009}, there is no distinction between the different rotation types. Both the radial pressure gradient and the radial electric field contribute to the subsonic toroidal rotation. The angular frequency of the toroidal rotation $\Omega_\zeta$ has two pieces: one driven by the radial electric field, $\Omega_{\zeta,E}=-\left({c}/{RB_\theta}\right)\left({\partial \phi_0}/{\partial r}\right)$, which corresponds to $E\times B$ plasma flow; and the other driven by the pressure gradient, $\Omega_{\zeta, p}=-\left({c}/{Zen_iRB_\theta}\right)\left({\partial p_i}/{\partial r}\right)$, corresponding to the diamagnetic flow. Here, $c$ is the speed of light, $R$ is the major radius, $B_\theta$ is the poloidal magnetic field, $\phi_0(r)$ is the long wavelength electrostatic potential, $r$ is the radial coordinate, $Ze$ is the ion charge, $n_i(r)$ is the ion density, and $p_i(r)$ is the ion pressure. The turbulent momentum pinch due to the radial electric field driven rotation is the only momentum pinch that has been investigated by other authors \cite{Peeters:POP2009}. In this Letter, the turbulent momentum pinch for the toroidal rotation driven by the radial pressure gradient, $\Omega_{\zeta, p}$, is calculated, and it is compared with the pinch for the rotation driven by the radial electric field, $\Omega_{\zeta, E}$. 
 
The two types of toroidal rotation have different origins. The radial electric field is related to the toroidal rotation via the Lorentz force, and as a result the radial electric field changes on the momentum confinement time scale \cite{Parra:NF2011}. Conversely, the pressure gradient is determined by the turbulent transport of energy and particles, changing on the energy confinement time scale.


We found that the effect of the two different types of rotation on the turbulent momentum transport are also different. The main difference is that the energy of the particles changes due to the radial electric field but not due to the radial pressure gradient. This is because the orbits of the particles in a tokamak deviate radially from the flux surface in which the magnetic field lies due to their magnetic drifts. The work on the particle done by the radial electric field is poloidally antisymmetric for an up-down symmetric tokamak, and it can give a nonzero momentum flux because it breaks the up-down symmetry of the turbulence. However, the pressure gradient cannot change the particle orbit or energy.

The dependence of the turbulent momentum pinch on the rotation type reveals a new symmetry breaking mechanism in the absence of flow in which the two different types of rotation are equal in size and opposite in sign. This intrinsic rotation generation mechanism can be important in the pedestal, a region of strong gradients observed in the edge of tokamaks in certain regimes \cite{Groebner:POP1998}, where both a large pressure gradient and a strong radial electric field contribute to the toroidal rotation in opposite directions. 

\emph{Radial momentum flux.} If there is no external momentum source, the radial transport of ion toroidal angular momentum, $\Pi$, determines the evolution of the toroidal flow, 
\begin{eqnarray}
\frac{\partial \left \langle n_i m_iR^2 \Omega_{\zeta}\right \rangle_{s} }{\partial t}=-\frac{1}{J}\frac{\partial (J \Pi)}{\partial r},
\end{eqnarray}
where $m_i$ is the ion mass, $J$ is the Jacobian and $\langle ...\rangle_{s}$ is the flux surface average. In steady state, the radial momentum flux has to be zero at every flux surface. The balance between non-zero pieces in $\Pi$ gives the radial profile of toroidal flow; e.g., $\Pi(\Omega_{\zeta}, \partial\Omega_{\zeta}/\partial r, ...)=0$ determines $\Omega_{\zeta}(r)$. The radial momentum flux is dominantly caused by the $E\times B$ radial drift due to the short wavelength fluctuating potential carrying the fluctuating toroidal angular momentum,
\begin{eqnarray}
\Pi&\simeq& \left\langle\left\langle(n_i m_iRV_{i,\zeta}^{tb}) (\mathbf{v}^{tb}_{E\times B}\cdot \mathbf{\hat{r}})  \right\rangle_s \right\rangle_T\\&=& \left\langle\left\langle \frac{-m_ic}{B}R\int d^3v f_{tb} (\mathbf{v}\cdot \mathbf{\hat{\zeta}}) (\nabla\phi^{tb} \cdot \mathbf{\hat{y}})  \right\rangle_s \right\rangle_T,
\end{eqnarray}
where $\mathbf{\hat{\zeta}}$ and $\mathbf{\hat{r}}$ are the toroidal and radial unit vectors, respectively, $B$ and $\mathbf{\hat{b}}$ are the magnitude and the unit vector of the magnetic field, respectively, $\mathbf{\hat{y}}=\mathbf{\hat{b}} \times  \mathbf{\hat{r}}$, and $\langle ...\rangle_{T}$ is a time average over several turbulence characteristic times. Here, $\mathbf{v}^{tb}_{E\times B}$ is the $E\times B$ drift due to the fluctuating electrostatic potential and $V_{i,\zeta}^{tb}$ is the fluctuating toroidal velocity. The fluctuating distribution function $f_{tb}$ and the fluctuating piece of the potential $\phi^{tb}$ due to microturbulence can be obtained from gyrokinetics. 

\emph{Gyrokinetics.}  
Microturbulence in a tokamak is well described by the gyrokinetic approximation. Gyrokinetics assumes that the gyromotion is much faster than the turbulent fluctuations ($\omega \ll \Omega_i$), and the turbulence characteristic length scales in the directions parallel and perpendicular to the static magnetic field are of the order of the ion Larmor radius and the size of the device, respectively ($\rho_i/l_\perp \sim 1$ and ${l_\perp}/l_\| \sim \rho_i/a \equiv \rho_{\star} \ll 1$), where $\omega$ is the frequency of turbulence, $\Omega_i$ is the ion Larmor frequency, $\rho_i$ is the ion Larmor radius, $a$ is the minor radius, and $l_\|$ and $l_\perp$ are the parallel and perpendicular turbulence length scale, respectively \cite{Catto:PP1978,Frieman:PF1982}. The dependence of the distribution function on the angle of Larmor gyration is eliminated by averaging over the fast Larmor gyration. As a result, the velocity space is described by two variables  (the kinetic energy $E={v^2}/{2}$ and magnetic moment $\mu={v_\perp^2}/{B}$, where $v_\perp=\sqrt{v^2-v_\|^2}$ and  $v_\|$ are the velocity perpendicular and parallel to the static magnetic field, respectively). After averaging over the gyromotion, the particle motion can be described by following the center of the Larmor gyration.

We assume that the ion distribution function $f_i=f_0^{(L)}+f^{(L)}_{tb}$ can be divided into two pieces: $f_0^{(L)}={n_i}/\left({\pi^{3/2}v_{ti}^{3}}\right)\exp\left(-{|\mathbf{v}-R\Omega_\zeta \hat{\zeta}|^2}/{v_{ti}^2}\right)\simeq f_{M}\left(1+\left({mv_\|}/{T_i}\right)\left({I\Omega_\zeta}/{B}\right)\right)$ is the lowest order non-fluctuating piece that includes the toroidal flow, and $f^{(L)}_{tb}$ is the fluctuating short wavelength piece whose size is $O(\rho_{\star}f_M)$. Here, $f_{M}$ is the Maxwellian distribution function, $I=B_\zeta R$, $B_\zeta$ is the toroidal magnetic field and $v_{ti}=\sqrt{2T_i/m_i}$ is the ion thermal velocity for the temperature $T_i$. We assume $B_\theta/B \ll 1$ and that the toroidal flow is sufficiently small to make $(B/B_\theta)\rho_\star v_{ti} \sim \Omega_{\zeta,p}R \sim \Omega_{\zeta,E}R \ll v_{ti}$. Similarly, the electrostatic potential can be divided into two pieces  $\phi=\phi_0+\phi^{tb}$. The gyrokinetic equation for the turbulent piece of the ion distribution function in the lab frame in the presence of the ion rotation due to both radial electric field and pressure gradient ($\Omega_\zeta=\Omega_{\zeta,E}+\Omega_{\zeta,p}$) is
\begin{eqnarray}
&\frac{\partial f_{tb}^{(L)}}{\partial t} +\left(v_\|\hat{b}+\mathbf{v}_M -\frac{c}{B} \nabla(\phi_0+\langle\phi^{tb} \rangle) \times \hat{b}\right) \cdot  \nabla f_{tb}^{(L)}\nonumber\\ & -\frac{Ze}{m_i}\mathbf{v}_M \cdot\nabla \phi_0  \frac{\partial f_{tb}^{(L)}}{\partial E} =\frac{c}{B} \nabla\langle\phi^{tb} \rangle \times \hat{b} \cdot  \nabla f_{0}^{(L)} \nonumber \\ &+ \frac{Ze}{m_i}[  v_\| \hat{b}+\mathbf{v}_M] \cdot\nabla\langle\phi^{tb} \rangle\frac{\partial f_{0}^{(L)}}{\partial E}+ \langle C(f_i) \rangle \label{GKE1}.
\end{eqnarray}
Here, $\mathbf{v}_M = \left({\mu}/{\Omega_i}\right)\hat{b} \times \nabla B+\left(v_\|^2/{\Omega_i}\right)\hat{b} \times (\hat{b} \cdot \nabla b)$ is the $\nabla B$ and the curvature drift, $\langle ...\rangle$ is the average over the gyromotion, and $C$ is the ion collision operator. Equation (\ref{GKE1}) only includes the terms related to the diamagnetic flow among the higher order corrections of order $({B}/{B_\theta})\rho_\star^2$. Other terms of order $({B}/{B_\theta})\rho_\star^2$ will be treated in a future publication \cite{Parra:NF2011,Parra:POP2012}. Notice that the acceleration and the drift of a single particle are not affected by the pressure gradient but are affected by the radial electric field $-\nabla \phi_0$, while the background distribution function is modified by both the pressure gradient and the radial electric field. The gyrokinetic equations can be derived for every species and they are coupled by imposing the quasineutrality condition.


The gyrokinetic equation with nonzero rotation can be described either in a lab frame (L) or a rotating frame (R). These frames are related by a rotating frame transformation, $ v_\|^\prime=v_\|-I\Omega_\zeta/B$ and $\varepsilon=E-I\Omega_\zeta v_\|/B$ \cite{Parra:NF2011}, where $v_\|^\prime$ and $\varepsilon$ are the parallel velocity and the kinetic energy variable in the rotating frame. In the rotating frame, the ion distribution function is $f_i=f_0^{(R)}(v_\|^\prime)+f^{(R)}_{tb}(v_\|^\prime)$ where the lowest order non-fluctuating piece is related to that in the lab frame by $f_0^{(R)}(v_\|^\prime)=f_{M}(v_\|^\prime)=f_0^{(L)}(v_\|)$, and the first order fluctuating piece is given by $f^{(R)}_{tb}(v_\|^\prime)=f^{(L)}_{tb}(v_\|)\simeq f^{(L)}_{tb}(v_\|^\prime)+\left({I\Omega_\zeta}/{B}\right)\left({\partial f^{(L)}_{tb}}/{\partial v_\|^\prime}\right)$.
Applying this transformation to Eq. (\ref{GKE1}), the gyrokinetic equation in the rotating frame is obtained,
\begin{eqnarray}
&\left(\frac{\partial }{\partial t}+ \Omega_\zeta R \hat{\zeta}\cdot \nabla \right)f_{tb}^{(R)}+ \big(v_\|^\prime \hat{b}-\frac{1}{n_im_i\Omega_i}\frac{\partial p_i}{\partial r} \hat{b} \times \nabla r \nonumber \\&+\mathbf{v}_M^\prime+\mathbf{v}_C^\prime- \frac{c}{B} \nabla\langle\phi^{tb} \rangle\times \hat{b} \big) \cdot \nabla f_{tb}^{(R)}\nonumber\\&- \frac{Ze}{m_i}\mathbf{v}_M^\prime \cdot\nabla r\left(-\frac{1}{Zen_i}\frac{\partial p_i}{\partial r}\right)  \frac{\partial f_{tb}^{(R)}}{\partial \varepsilon} =\frac{c}{B} \nabla\langle\phi^{tb} \rangle \times \hat{b} \cdot  \nabla f_{0}^{(R)}\nonumber\\&+ \frac{Ze}{m_i}[  v_\|^\prime\hat{b}+\mathbf{v}_M^\prime+\mathbf{v}_C^\prime]  \cdot\nabla\langle\phi^{tb} \rangle\frac{\partial f_{0}^{(R)}}{\partial \varepsilon}+ \langle C(f_i) \rangle \label{GKE2},
\end{eqnarray}
where $\mathbf{v}_M^\prime= \left({\mu}/{\Omega_i}\right)\hat{b} \times \nabla B+\left(v_\|^{\prime2}/{\Omega_i}\right)\hat{b} \times (\hat{b} \cdot \nabla b)$ is the $\nabla B$ and the curvature drift in the rotating frame, and $\mathbf{v}_C^\prime=\left({2v_\|^\prime\Omega_\zeta}/{\Omega_i}\right)\hat{b} \times [(\nabla{R} \times \hat{\zeta})\times \hat{b}]$ is the drift due to the Coriolis force. 

In the frame of rotation driven only by the radial electric field ($\Omega_{\zeta,E}\neq 0$ and $\Omega_{\zeta,p}=0$), the gyrokinetic equation is modified only by the additional Coriolis terms \cite{Peeters:POP2009}, without any energy derivative of the turbulent distribution function. Conversely, the rotation driven by the pressure gradient ($\Omega_{\zeta,E}=0$ and $\Omega_{\zeta,p}\neq0$) does not include the energy derivative terms in the lab frame but it does have these terms in the rotating frame. This additional acceleration term that depends on the type of the rotation,  $\left({ZeRB_\theta}/{m_ic}\right)\mathbf{v}_M^\prime \cdot\nabla r\left({\partial f_{tb}}/{\partial  \varepsilon}\right)\Omega_{\zeta,p}$, results in different momentum pinches for the two rotations in either lab frame (\ref{GKE1}) or rotating frame (\ref{GKE2}). The acceleration term breaks the symmetry of the turbulence by introducing an up-down asymmetry effect in the poloidal coordinate ($\theta$). For example, for a tokamak with circular flux surfaces, the term is proportional to $\sin\theta$ which is odd in $\theta$. The other rotation type dependent term, $ \left({RB_\theta}/{B}\right) \nabla r\times \hat{b} \cdot  \nabla f_{tb}\Omega_{\zeta,p}$, only gives a Doppler shift to the fluctuation that cannot change the radial transport. 

\emph{Numerical results.}   
To model the momentum transport $\Pi(\Omega_{\zeta}, \partial\Omega_{\zeta}/\partial r)$, we assume that $\Omega_\zeta$ is sufficiently small and linearize around $\Omega_\zeta=0$ and $\partial\Omega_{\zeta}/\partial r=0$, giving
 \begin{eqnarray}
\Pi=\Pi_{int}-P_{\zeta} m_i \langle R^2\rangle_s \Omega_{\zeta}- \chi_{\zeta} m_i\langle R^2\rangle_s\frac{\partial\Omega_{\zeta}}{\partial r}, \label{Pi0}
\end{eqnarray}
 where $\Pi_{int}$ is the intrinsic toroidal angular momentum flux in the absence of rotation and rotation shear ($\Omega_\zeta=0$ and ${\partial \Omega_\zeta}/{\partial r}=0$), $P_{\zeta}$ is the pinch coefficient, and $\chi_\zeta$ is the toroidal momentum diffusivity. We use the gyrokinetic code GS2 \cite{Dorland:PRL2000} to evaluate the diffusion and pinch coefficients for the different types of rotation. In this Letter, we assume that the momentum diffusivity coefficients are not different for the different type of rotations and $\chi_\zeta$ is inferred from a fixed Prandtl number, the ratio of the momentum diffusivity to the ion heat flux diffusivity, ($Pr\equiv {\chi_\zeta}/{\chi_i}=-\left({\chi_\zeta}/{Q_i}\right)\left({n_i \partial T_i}/{\partial r}\right)\simeq 0.5$), obtained in a simulation with velocity shear. 
 

The numerical results for the momentum pinches are given in Fig. 1. The default plasma parameters for the simulations are $R_0/L_{T}=9.0$, $R_0/L_{n}=9.0$, $q=2.5$, $r/a=0.8$, $R_0/a=3.0$, and $\hat{s}=0.8$ where $L_{T}=-T_i/(dT_i/dr)$ and $L_{n}=-n_i/(dn_i/dr)$ are characteristic lengths of temperature and density, respectively, $R_0\simeq\langle R \rangle_s$, $q$ is the safety factor, and $\hat{s}$ is the magnetic shear. We use the rotation peaking factor ${P_{\zeta}}/{\chi_\zeta}$ to quantify the strength of the pinch (${P_{\zeta}}/{\chi_\zeta}=-{\Omega_\zeta}^{-1}\left({\partial\Omega_\zeta}/{\partial r}\right)$ is the rotation peaking for $\Pi=0$ and $\Pi_{int}=0$ in Eq. (\ref{Pi0})). Fig. 1 shows that the rotation peaking factors are ${P_{\zeta,E}}/{\chi_\zeta}\simeq 2.9/R_0$ for $\Omega_{\zeta,E}$, and ${P_{\zeta,p}}/{\chi_\zeta}\simeq 3.5/R_0$ for $\Omega_{\zeta,p}$. The difference is about $22\%$ of the pinch. 

Interestingly, the difference of the pinches results in negative momentum flux even for zero rotation ($\Omega_{\zeta}=\Omega_{\zeta,E}+\Omega_{\zeta,p}=0$, and $\Omega_{\zeta,p}>0$ because the pressure decreases with radius). The inward intrinsic momentum flux in the absence of rotation ($\Pi_{int}<0$) results in rotation peaking. The toroidal rotation peaking factor, $\left({a}/{\Omega_{\zeta,p}}\right)\left({\partial\Omega_{\zeta}}/{\partial r}\right)=\left(a/\Omega_{\zeta,p}\right)\left({\Pi_{int}}/\chi_\zeta\right)\left(1/m_i \langle R^2\rangle_s \right)$, as a result of this intrinsic momentum transport can be calculated by assuming $\Pi=0$ and $\Omega_\zeta=0$ in Eq. (\ref{Pi0}).
Fig. 2 shows the toroidal rotation peaking factor with ${\Omega_{\zeta,p}R_0}/{v_{ti}}=0.3$ and ${\Omega_{\zeta,E}R_0}/{v_{ti}}=-0.3$. Assuming that this ${\partial\Omega_{\zeta}}/{\partial r}$ is a good estimate for the typical size of the rotation gradient, it results in a peak flow in the core of up to $40 \%$ of the piece of the rotation due to the negative radial pressure gradient in the pedestal. 
\begin{figure}
\includegraphics[scale=0.4]{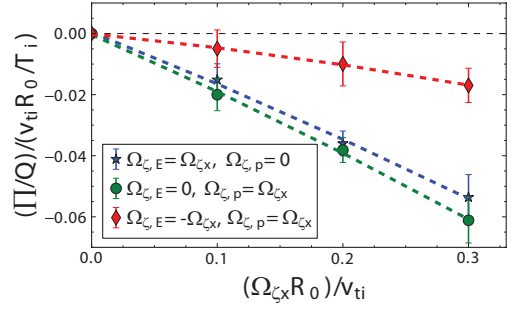}
\caption{Time averaged ratio of ion toroidal angular momentum flux ($\Pi$) to ion heat flux ($Q_i$) as a function of rotation (${\Omega_{\zeta x}R_0}/{v_{ti}}$) for zero rotation shear and the different types of rotation: radial electric field driven rotation (blue-star), pressure gradient driven rotation (green-circle) and opposite rotations of the two types (red-diamond). The error bars show the standard deviation of the fluxes from the time average values due to the typical turbulent fluctuations.} 
\end{figure}

\begin{figure}
\includegraphics[scale=0.36]{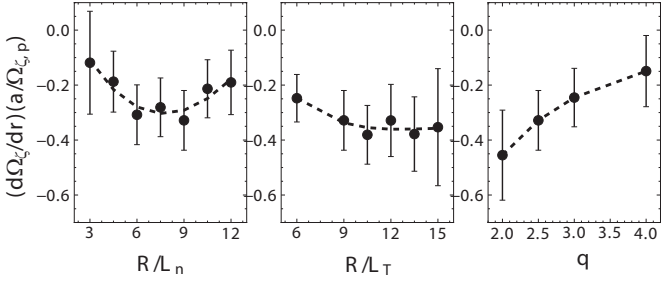}
\caption{The normalized rotation peaking factor generated by the intrinsic momentum fluxes in terms of (a)  the density gradient ($R/L_{n}$), (b) the temperature gradient ($R/L_{T}$), and (c) the safety factor ($q$). The parameters except the scanned variable are the same as in Fig 1.}
\end{figure}

The existence of two different pinches is one of the sources of intrinsic rotation generation according to the following restatement of Eq. (\ref{Pi0}),
 \begin{eqnarray}
\Pi&=&\bigg[\Pi^\prime_{int}-\left(\frac{P_{\zeta,E}-P_{\zeta,p}}{2}\right)m_i \langle R^2\rangle_s(\Omega_{\zeta,E}-\Omega_{\zeta,p})\bigg]\nonumber\\&&-\bigg[\left(\frac{P_{\zeta,E}+P_{\zeta,p}}{2}\right)m_i \langle R^2\rangle_s \left(\Omega_{\zeta,E}+\Omega_{\zeta,p}\right)\bigg]\nonumber\\&&-\chi_{\zeta} m_i\langle R^2\rangle_s \frac{\partial\Omega_{\zeta}}{\partial r},
\end{eqnarray}
 where $\Pi_{int}=\Pi^\prime_{int}-\left(\frac{P_{\zeta,E}-P_{\zeta,p}}{2}\right)m_i \langle R^2\rangle_s(\Omega_{\zeta,E}-\Omega_{\zeta,p})$ and $\Pi^\prime_{int}$ is the intrinsic toroidal angular momentum flux for $\Omega_{\zeta,E}=0$, $\Omega_{\zeta,p}=0$, and  $\partial\Omega_\zeta/\partial r=0$. 

\emph{Result analysis.} 
  The toroidal momentum flux can be analyzed using the symmetry of the turbulence in $\theta$, $v_\|^\prime$ and $k_r$ \cite{Parra:POP2011}, where $k_r$ is the radial wavenumber of the turbulence. The momentum flux in the rotating frame can be written as $\Pi^{(R)}=\sum_{k_r}\int d\theta dv_\|^\prime \pi(\theta,v_\|^\prime,k_r)$, and  $\pi$ can be expanded in $\rho_\star$ and  $(B/B_\theta)\rho_\star$: $\pi=\pi_1+\pi_2+...$ where $\pi_1\propto f_{tb,1} (\mathbf{v^\prime}\cdot \mathbf{\hat{\zeta}}) (\nabla\phi_1^{tb} \cdot \mathbf{\hat{y}})$ and $\pi_2\propto f_{tb,2} (\mathbf{v^\prime}\cdot \mathbf{\hat{\zeta}}) (\nabla\phi_1^{tb} \cdot \mathbf{\hat{y}})+f_{tb,1} (\mathbf{v^\prime}\cdot \mathbf{\hat{\zeta}}) (\nabla\phi_2^{tb} \cdot \mathbf{\hat{y}})\sim (B/B_\theta)\rho_\star\pi_1$. Because the first order fluctuating ion distribution function, $f_{tb,1}\sim  O(\rho_{\star}f_M)$, and the corresponding potential $\phi_1^{tb}$ have the parity $f_{tb,1} (\theta,v_\|^\prime,k_r)= -f_{tb,1} (-\theta,-v_\|^\prime,-k_r)$ and $\phi_1^{tb} (\theta,k_r)=-\phi_1^{tb} (-\theta,-k_r)$, the toroidal momentum flux to this order is antisymmetric, $\pi_1^{tb}(\theta,v_\|^\prime,k_r)= -\pi_1^{tb} (-\theta,-v_\|^\prime,-k_r)$. As a result, the total sum over $\pi_1$ vanishes.

The different parity of the distribution function $f_{tb,2}$ and the potential $\phi_2^{tb}$  (i.e. $f_{tb,2}(\theta,v_\|^\prime,k_r)= f_{tb,2}(-\theta,-v_\|^\prime,-k_r)$ and $\phi_2^{tb} (\theta,k_r)= \phi_2^{tb} (-\theta,-k_r)$) are caused by the Coriolis terms breaking the symmetry in $v_\|^\prime$ and the additional acceleration term, $\left({ZeRB_\theta}/{m_ic}\right)\mathbf{v}_M^{\prime} \cdot\nabla r\left({\partial f_{tb}}/{\partial  \varepsilon}\right)\Omega_{\zeta,p}$, breaking the symmetry in $\theta$. We focus on the additional acceleration term for the case with $\Omega_{\zeta,E}+\Omega_{\zeta,p}=0$ because the Coriolis term has already been studied \cite{Peeters:PRL2007}. Using a balance between the acceleration term, $\left({ZeRB_\theta}/{m_ic}\right)\mathbf{v}_M^{\prime} \cdot\nabla r\left({\partial f_{tb,1}}/{\partial  \varepsilon}\right)\Omega_{\zeta,p}$ and the time derivative of $f_{tb,2}$ (${\partial f_{tb,2}}/{\partial t}\sim {f_{tb,2}}/{\tau_{nl}}$, with $\tau_{nl}$ the nonlinear decorrelation time  \cite{Barnes:PRL2011}), we obtain
\begin{eqnarray}
f_{tb,2} \sim \tau_{nl} \frac{Ze}{T_i}(v_\|^{\prime2}+v_\perp^{\prime2}/2)\Omega_{\zeta,p}\frac{B_\theta}{B}\sin \theta\frac{\partial f_{tb,1}}{\partial \varepsilon},\label{g2}
\end{eqnarray}
where we have used the magnetic drift of concentric circular flux surfaces to make the dependence on $\theta$ more transparent. The higher order pieces give the even parity of the next order contribution to the momentum flux: $\pi_2(\theta,v_\|^\prime,k_r)= \pi_2 (-\theta,-v_\|^\prime,-k_r)$. Accordingly, the ratio of the momentum flux to the heat flux decreases when the safety factor increases because of $\left(B_\theta/B\right) \simeq (r/R)(1/q)$ ($\tau_{nl}$ is barely modified by the change of the safety factor in our simulations). More analysis is needed to completely understand the dependence on $R/L_n$ and $R/L_T$ in Fig. 2. 


\begin{figure}
\includegraphics[scale=0.35]{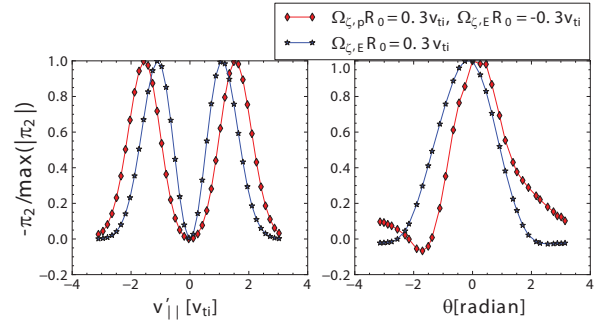}
\caption{The normalized profiles of the even parity momentum flux : (a) $\int_{\theta\text{min}}^{\theta\text{max}} {d\theta}\pi_2 $ in terms of $(v_\|^\prime/v_{ti})$, and (b) $\int_{0}^{\infty} dv_\|^\prime \pi_2 $ in terms of $\theta$.}
\end{figure}
 Fig. 3 shows the intrinsic momentum flux as a function of $v_\|^\prime$ and $\theta$ for $\left({\Omega_{\zeta,p}R}/{v_{ti}}\right)=0.3$ and $\left({\Omega_{\zeta,E}R}/{v_{ti}}\right)=-0.3$. The dependence on $v_\|^\prime$ and $\theta$ are different from those for the Coriolis momentum pinch due to the radial electric field by itself. The difference is due to the different mechanisms for the symmetry breaking. The profile in $v_\|^\prime$ in Fig 3-(a) shows that the intrinsic momentum transport has a larger contribution from particles with the large parallel velocity because $f_{tb,2}$ has a higher order moment in $v_\|^\prime$ due to the magnetic drift effect on the acceleration. Physically, the particles with large parallel velocity have wider orbits than slow particles and they exchange more energy with the radial electric field. The asymmetric $\theta-$profile in Fig 3-(b) can be explained by the factor $\sin\theta$ in Eq. (\ref{g2}), which is odd in $\theta$. 
 
\emph{Discussion.} 
The intrinsic momentum flux due to the two opposite pieces of the rotation in the pedestal can be one of the origins for intrinsic toroidal rotation. When ions are not rotating toroidally in the pedestal, a negative radial electric field must exist to balance the radial pressure drop ($\Omega_{\zeta,E} \sim -\Omega_{\zeta,p}<0$). The intrinsic momentum transport occurs due to the acceleration of the particles in the radial electric field that breaks the symmetry of the turbulence. For example, for $\Delta T_i \sim -1keV$ in $\Delta r \sim 1\text{cm}$ with $B_\theta\sim 0.5T$, $\Omega_{\zeta,p}R_0 \sim 200\text{km/s}$. The intrinsic momentum transport results in rotation peaking in the core whose size can be about $40\%$ of $\Omega_{\zeta,p}R_0$,  i.e., $\Omega_{\zeta}R_0 \sim 80\text{km/s}$.

This work was supported in part by Samsung scholarship, by U.S. DoE FES Postdoctoral Fellowship, and by US DoE Grant No. DE-SC008435. This research used computing resources of NERSC.





\end{document}